\documentclass[aps,prd,reprint,round]{revtex4-2}
\usepackage[T1]{fontenc}
\usepackage[utf8]{inputenc}
\usepackage{ae,aecompl} % MNRAS-recommended font support

\usepackage{amsmath,amssymb}
\usepackage{bm}
\usepackage{physics} % acceptable, but be conservative

\usepackage{graphicx}
\usepackage{caption}

\usepackage{subcaption}

\usepackage{xcolor}
\definecolor{linkcolor}{rgb}{0.0,0.3,0.5}

\usepackage[hypertexnames=false, unicode, colorlinks=true, linkcolor=linkcolor,
citecolor=linkcolor, filecolor=linkcolor,urlcolor=linkcolor,
pdfusetitle]{hyperref}

\usepackage{xspace}
\usepackage{microtype}
\usepackage{import}

\newcommand{\beq}{\begin{equation}}
\newcommand{\eeq}{\end{equation}}

\def\k{{\boldsymbol{k}}}

\def\r{{\boldsymbol{r}}}
\def\v{{\boldsymbol{v}}}

\def\VEV#1{\left\langle #1 \right\rangle}

\newcommand{\be}{\begin{equation}}
\newcommand{\ee}{\end{equation}}
\newcommand{\ba}{\begin{eqnarray}}
\newcommand{\ea}{\end{eqnarray}}

\def\fun#1#2{\lower3.6pt\vbox{\baselineskip0pt\lineskip.9pt
        \ialign{$\mathsurround=0pt#1\hfill##\hfil$\crcr#2\crcr\sim\crcr}}}

\definecolor{darkgreen}{RGB}{1,150,24}

\begin{document}
\title[Short title]{Forecasting neutrino mass constraints\\ from the \textit{Nancy Grace Roman Space Telescope}}

\author{Francesco Spezzati}
\email{spezzati@caltech.edu}
\affiliation{IPAC, Caltech, 1200 E California Blvd, 91125 Pasadena, CA}
\author{Yun Wang}
\email{wang@ipac.caltech.edu}
\affiliation{IPAC, Caltech, 1200 E California Blvd, 91125 Pasadena, CA}
\author{Andrew Hearin}
\email{ahearin@anl.gov}
\affiliation{HEP Division, Argonne National Laboratory, 9700 South Cass Avenue, Lemont, IL 60439, USA}

\begin{abstract}
We present realistic forecasts for the constraining power of the Nancy Grace Roman Space Telescope on fundamental cosmological parameters, with particular emphasis on the absolute neutrino mass scale, using full-shape analyzes of the galaxy power spectrum. We analyze simulated lightcone mock catalogs of H$\alpha$ emission-line galaxies spanning the redshift range $0.5 < z < 2$ over $2400\ \mathrm{deg}^2$, designed to reproduce the expected properties of the Roman High Latitude Wide Area Spectroscopic Survey. We perform parameter inference on the galaxy power spectrum multipoles using two complementary theoretical frameworks: a model-dependent approach based on the Effective Field Theory of Large-Scale Structure (EFT of LSS) within $\Lambda$CDM, and a model-independent phenomenological approach that makes no assumptions about the background cosmological model. In the $\Lambda$CDM analysis, we find $m_\nu < 0.380(0.162)\ \mathrm{eV}$ at $95(68)$\% C.L. using Big Bang Nucleosynthesis (BBN) prior and a broad prior on $n_s$, which tightens to $m_\nu < 0.276(0.121)\ \mathrm{eV}$ when Planck priors on $\omega_b$, $\omega_\mathrm{cdm}$, and $n_s$ are added. Our forecasts show that Roman can additionally constrain $H_0$,
$\Omega_m$, and $\sigma_8$ with precisions of $1.3\%$, $4.3\%$, and $2.9\%$ in line with Stage IV galaxy survey measurements and forecasts. In the model-independent analysis, we demonstrate that the phenomenological model can robustly recover unbiased measurements of the angular diameter distance, the Hubble parameter, and the growth of structure across all redshift bins, in the same range of scales as the EFT model, and obtain $m_\nu < 0.63(0.36)\ \mathrm{eV}$ at $95(68)$\% C.L.\ when Planck priors are included.
\end{abstract}

\maketitle

\section{Introduction}
The study of the large-scale structure (LSS) of the Universe is entering a golden era. A new generation of galaxy surveys—including DESI~\cite{DESI:2016fyo} and Euclid~\cite{Euclid:2024yrr},
%and SPHEREx~\cite{SPHEREx2014}
%\red{(Yun: I removed SPHEREx, which uses $\sigma_z/(1+z)=0.1$ in their cosmological analysis, which is the photo-z regime. It does intensity mapping essentially, and does not map the 3D LSS.)}
—is mapping the three-dimensional distribution of matter over unprecedented cosmic volumes. By measuring the positions and redshifts of tens of millions of galaxies with exquisite precision, these experiments are producing the most detailed reconstructions of large-scale structure (LSS) to date. Such datasets enable high-precision measurements of galaxy clustering across a wide range of scales and redshifts, offering a powerful avenue to constrain the fundamental parameters governing the composition, geometry, and evolution of the Universe.

Within this rapidly evolving landscape, the Nancy Grace Roman Space Telescope (hereafter Roman)~\cite{spergel2015wide} is expected to play a key role in the coming years. Through its High Latitude Wide Area Spectroscopic Survey (HLWASS)~\cite{Wang:2021oec}, Roman will measure the positions of approximately $20$ million galaxies in the redshift range $0.5<z<2$ using H$\alpha$, and 2 million galaxies at $2<z<3$ using [OIII], over an area of $\sim2400 \ \mathrm{deg}^2$. This combination of wide sky coverage, high number density, and broad redshift reach will enable exceptionally precise galaxy clustering measurements at intermediate and high redshifts, complementing and extending the scientific reach of other current and forthcoming surveys.

Among the key open questions in fundamental physics that can be addressed with LSS data, the absolute mass scale of neutrinos occupies a central position. Neutrino oscillation experiments have conclusively demonstrated that at least two neutrino species are massive, giving $\sum_im_i\gtrsim0.06(0.10)\ \mathrm{eV}$ for normal (inverted) hierarchy~\cite{Super-Kamiokande:1998kpq,SNO:2001kpb,SNO:2002tuh}; yet, such experiments are insensitive to the absolute neutrino mass scale. Cosmology provides a complementary and, in principle, uniquely sensitive approach to this problem. Because neutrinos remain relativistic well after decoupling and subsequently transition to non-relativistic behavior, their finite masses leave characteristic signatures in the evolution of density perturbations. In particular, massive neutrinos suppress matter perturbations and the growth of structure below their free-streaming scale, introducing a scale- and redshift-dependent modification of galaxy clustering observables. These effects make the large-scale distribution of galaxies a powerful probe of neutrino physics. For updated cosmological constraints on the sum of neutrino masses, see e.g.,~\cite{DESI:2025zgx,Elbers:2025vlz}.

In this work, we forecast the ability of the Roman telescope to constrain fundamental cosmological parameters using measurements of the galaxy power spectrum at $0.5<z<2$, with particular attention to its sensitivity to neutrino mass effects. To achieve this in a realistic and robust manner, we analyze a realistically simulated Roman galaxy mock catalog designed to reproduce the expected properties of Roman observations in terms of redshift distribution and galaxy number density. By performing parameter inference on this galaxy mock catalog the same way we would on real Roman data, we assess the constraining power of Roman under realistic observational and modeling assumptions, and we quantify its expected contribution to future precision studies of cosmology and fundamental physics.
\section{Theoretical models}
The galaxy power spectrum is modeled using two complementary approaches. In the first (hereafter the \textit{model-dependent} template), the theoretical two-point correlator is constructed under the assumption that the Universe is well described by the standard $\Lambda$CDM cosmology. This framework enables a self-consistent treatment of the late-time evolution of the galaxy overdensity field, extending into the mildly non-linear regime, and allows cosmological inference to be performed directly on the fundamental parameters of the $\Lambda$CDM model.

We also consider a \textit{model-independent} approach in which the power spectrum template is built without assuming a specific underlying cosmological model. This strategy permits the extraction of cosmological information from galaxy clustering data with minimal theoretical priors. We compare these two pipelines and quantify their respective strengths and limitations in the context of a realistic galaxy clustering analysis.
\subsection{$\Lambda\mathrm{CDM}$}
In this work, the $\Lambda$CDM analysis is carried out within the framework of the Effective Field Theory of Large Scale Structure (EFT of LSS)~\cite{Baumann:2010tm,Carrasco:2012cv}. For recent full shape power spectrum analyzes of galaxy clustering data employing this formalism, see e.g.,~\cite{DAmico:2019fhj,Zhang:2021yna,Ivanov:2019hqk,Ivanov:2019pdj,Chudaykin:2025aux}.
Within this approach, the perturbative solutions of the equations of motion for the matter overdensity and velocity fields are assumed to be separable in time and space. This assumption is exact in an Einstein–de Sitter universe and remains an excellent approximation in a \(\Lambda\)CDM cosmology with massless neutrinos~\cite{Bernardeau:2001qr}. However, it is, in principle, no longer valid in the presence of massive neutrinos. In fact, the scale-dependent suppression of structure growth induced by massive neutrinos breaks down the time–space separability of the solutions . On scales smaller than the neutrino free-streaming scale, their large thermal velocities prevent efficient clustering, while on scales larger than the free-streaming scale, neutrinos behave as cold dark matter and contribute to gravitational collapse (see e.g.,~\cite{Lesgourgues:2006nd} for an extensive review on the subject). As a consequence, the growth of matter perturbations becomes scale dependent, which formally requires the computation of modified perturbation theory kernels that consistently account for massive neutrinos at the non-linear level~\cite{Blas:2014hya,Garny:2020ilv,Garny:2022fsh,Aviles:2021que,Noriega:2022nhf}. However, it has been shown that these effects lead to negligible corrections to the matter power spectrum at the level of precision achievable by current galaxy clustering measurements~\cite{Aviles:2021que,Maus:2024sbb,Garny:2020ilv}. For this reason, throughout this work, we compute the linear matter power spectrum including massive neutrinos, consistently modeling the scale dependence of the growth factor $D(k,t)$ and the growth rate $f=\frac{d\ln D(k,t)}{d \ln a}$, but we will employ standard perturbation theory kernels to evaluate the non-linear contributions to the two-point correlators.

Massive neutrinos modify both the $\textit{total}$ matter power spectrum and the $\mathrm{CDM}+$baryons power spectrum in a different way compared to a universe with $m_\nu=0$.
In the presence of massive neutrinos, the total matter overdensity is
\ba
  &&  \delta_m=f_\nu\delta_{\nu}+(1-f_\nu)\delta_{cb}, \\
&& f_{\nu}=\omega_\nu/\omega_m, \hspace{0.1in}\omega_m=\omega_b+\omega_c+\omega_\nu, \hspace{0.1in}\omega_\nu=\frac{m_{\nu,\mathrm{eV}}}{93.1}\nonumber
\ea
where $\delta_{cb}$ is the CDM+baryons perturbation and $m_{\nu,\mathrm{eV}}$ is the sum of the neutrino masses.  On scales larger than the neutrinos free streaming scale ($k\ll k_\mathrm{fs}$), $\delta_{cb}\sim\delta_\nu$, so $\delta_m\sim\delta_{cb}$ while for $k\gg k_\mathrm{ks}$, the matter perturbation is suppressed to $\delta_m\sim(1-f_\nu)\delta_{cb}$. Additionally, on scales smaller than the free streaming length, since neutrinos contribute to the background density but not to the clustering, the balance between the gravitational pull and the expansion rate in the equation of motion of $\delta_{cb}$ changes compared to the massless neutrino case. This effect suppresses the growth of $cb$ perturbations by roughly $\delta_{cb}\propto a^{1-\frac{3}{5}f_\nu}$~\cite{Bond:1980ha,Lesgourgues:2006nd} %\red{(Yun: need a reference here)}\FS{added}.
These two effects combined modify the $\textit{total}$ matter power spectrum. On the other hand, the suppression of the CDM+baryon power spectrum at scales smaller than $k_\mathrm{fs}$ is only due to the suppressed growth of  perturbations.%\red{(Yun: hanging incomplete sentence!)}\FS{fixed}

Since the $total$ matter power spectrum and the CDM+baryons ($cb$) power spectrum are modified differently by massive neutrinos, one needs to specify whether galaxies are biased tracers of the $total$ matter field or of the $cb$ field.
The equivalence principle implies that the bias operators in the perturbative bias expansion are constructed from double derivatives of the gravitational potential $\phi$~\cite{Desjacques:2016bnm}, which is sourced by the $total$ matter overdensity. Since the bias expansion is non-local in time, and since during most of the galaxy formation timescale neutrinos were not clustered at the scales of interest for a typical galaxy survey, $\delta_g\propto b_1\delta_{cb}(1-f_{\nu})$~\cite{Racco:2024lbu}.
This redefinition removes the leading order scale dependence of the linear bias observed when it is defined with respect to the total matter field~\cite{Raccanelli:2017kht,Villaescusa-Navarro:2013pva,Castorina:2013wga,Costanzi:2013bha}. The residual scale dependence of the linear bias was studied in~\cite{Chiang:2018laa,Chiang:2017vuk,LoVerde:2014rxa,LoVerde:2016ahu,LoVerde:2014pxa} and it is usually assumed to be negligible for the precision of Stage IV galaxy surveys. In practical LSS analysis, the overall factor $(1-f_\nu)$ is reabsorbed inside the unknown linear bias parameter so that we have $\delta_g\sim b_{1,\mathrm{eff}}\delta_m$, where $b_{\mathrm{eff}}$ is estimated from the data.
When dealing with redshift space distortions (RSD), \cite{Verdiani:2025znc} showed recently that assuming $\v_g=\v_{cb}$ (i.e., galaxies fall at the same rate as the $cb$ field) is a better approximation than the ansatz $\v_g=\v_{m}$ when compared to simulations. While a self consistent theoretical derivation of this result is still the subject of ongoing study, we adopt this ansatz in all that follows. We plan to investigate possible deviations from this approximation in a forthcoming work.

Keeping all these considerations in mind, the EFTofLSS model for the galaxy power spectrum multipoles used in this work can be written as~\cite{Ivanov:2019pdj,Chudaykin:2020aoj}:

\begin{align}
\label{eq:model_EFT}
        P_{g,\ell}(k,z)=\frac{(2\ell+1)}{2}&\int d\mu \big[P_{\ell,g,\mathrm{lin}}(\k,z)+P_{\mathrm{\ell,g,1-loop}}(\k,z)\\\nonumber&+ P_{\ell,\mathrm{ctr}}(\k,z)+P_{\ell,\mathrm{stoch.}}(\k,z)\big]\,.
\end{align}

The first and second terms on the right hand side of Equation~\eqref{eq:model_EFT} are the tree level matter power spectrum and the 1-loop power spectrum in Eulerian perturbation theory (EPT). The third term is introduced in the EFT approach to absorb the back-reaction of small non-linear scales into the large scale model. Explicitly, it reads:
\begin{equation}
    P_{\ell,\mathrm{ctr}}(\k,z)=c_{\ell}\mu^\ell f^{\ell/2} k^2P_{\ell,\mathrm{lin}}(\k,z)\,,
\end{equation}
where the functional form of each counterterm is dictated by the symmetries of the Universe, while their amplitudes have to be fitted directly from the data.
The presence of massive neutrinos does not require the addition of new counterterms for the precision of Stage IV galaxy surveys~\cite{Racco:2024lbu,Senatore:2017hyk}.

The stochastic noise due to galaxy formation is modeled as
\begin{equation}
    P_{\ell,\mathrm{stoch.}}(\k,z)=\frac{\epsilon_0}{\bar{n}}+\epsilon_2\mu^2k^2\,,
\end{equation}
where we have neglected the $\epsilon_1k^2$ term since, as shown for example in\cite{Schmittfull:2018yuk,Euclid:2026cpu,Chudaykin:2025aux}, its impact on the posterior parameters is negligible for the scales considered in this work.
The final ingredient for a consistent model for the EFT power spectrum is the IR resummation scheme, which accounts for the smearing of the BAO features due to large scale bulk flows that cannot be captured by EPT. Different resummation procedures have been proposed in the literature~\cite{Senatore:2017pbn,Senatore:2014via,Ivanov_IR,Baldauf_IR}, and they have been shown to produce almost identical results~\cite{Nishimichi:2020tvu,Maus:2024sbb}. We adopt the approach proposed in \cite{Senatore:2014via,Senatore:2017pbn}, which is implemented in the \texttt{PyBird} code~\cite{DAmico:2020kxu} used in this work to produce EFT predictions. We have modified the code to consistently include the presence of massive neutrinos, as described in the previous paragraphs.
We make use of the large scale limit of the growth rate $f(k\rightarrow0,z)\sim f (z)$ in the IR resummation formula, as the impact of the scale dependence of the growth rate on the damping of the BAO peak can be shown to be negligible for the scales considered here.

\subsection{Model-Independent Approach}
Most full-shape analyzes of galaxy two-point statistics rely on the assumption of a fiducial cosmological model to solve the equations governing the evolution of the galaxy overdensity and velocity fields. Departures from the standard $\Lambda\mathrm{CDM}$ paradigm are typically explored either by repeating the analysis for a discrete set of alternative models (see e.g.,~\cite{Taule:2024bot,Piga:2022mge}) or by introducing simple parameterizations —such as the $w_0w_a$ expansion~\cite{Chevallier:2000qy,Linder:2002et}—that re-map $\Lambda\mathrm{CDM}$ degrees of freedom but are not directly motivated by first principles.
An alternative and more agnostic approach is, however, possible. Galaxy clustering data can be analyzed without assuming a specific model for the late-time evolution of the Universe. In this framework, model-independent theoretical templates for two-point correlators are constructed up to mildly non-linear scales, and the cosmological information is compressed into a set of effective, model-independent parameters that can be robustly constrained by the data. This strategy ensures that the inference is not biased by an assumed background model and allows for cosmological model selection to be performed a posteriori, once the data have been analyzed in a fully consistent and unbiased manner.

To model the galaxy power spectrum in an agnostic, model-independent way, one can build the theoretical template based on phenomenological considerations and fitting formulae obtained from N-body simulations. Here, we adopt a phenomenological model for the galaxy power spectrum in redshift space based on the works of \cite{Wang:2014qoa,Wang:2016rmr,Zhai:2020yuc,McCarthy:2022qny} (for other model independent methods for the galaxy power spectrum, see e.g.,~\cite{DAmico:2021rdb,Marinucci:2024add,Peron:2025lgh,Amendola:2023awr,Brieden:2021edu,Brieden:2022ieb})

In this model, the galaxy power spectrum is given by
\begin{align}\label{eq:model_ind}
    P_g(k,\mu,z)=&A(z)[1+\mu^2W(k,z)\beta(z)]^2T_{dw}^2(k)k^{n_s}\\ \nonumber \times &e^{\frac{-(k\mu\sigma_v)^2}{2}}\left(\frac{1+Q(z)k^2}{1+C(z)k+\frac{Q(z)}{10}k^2}\right)\,,
\end{align}
where $A(z)$ and $\beta(z)$ encode amplitude and linear redshift space distortion information, respectively. Non-linearities in the biased field are modeled by a fitting polynomial in $k$~\cite{2dFGRS:2005yhx}. A Gaussian damping  and the kernel $W(z,k)$ are introduced phenomenologically to model nonlinear RSD~\cite{Percival:2008sh,1992MNRAS.259..494P,Zheng:2013ora}. As in \cite{Wang:2014qoa,Zhai:2019hjt,McCarthy:2022qny,Wang:2016rmr} we use a simple expression for $W(k,z)$:
\begin{equation}
    W(k,z)=\frac{1}{1+\Delta\alpha\Delta^2(k)}\,,
\end{equation}
where $\Delta \alpha$ is a $\mathcal{O}(1)$ free parameter of the theory, and $\Delta^2=\frac{k^3P(k)}{2\pi^2}$ is the dimensionless matter power spectrum related to the variance of the matter overdensity field in configuration space.
In this phenomenological model, the non-linear damping of the BAO due to bulk flows is accounted for in the wiggly part of the matter transfer function, and the damping factor is given by the expression~\cite{2013MNRAS.430.2446W}
\begin{equation}
    \Sigma^2_{\mathrm{pheno}}=(k^*)^{-2}[1-\mu^2+\mu^2(f^*(z)+1]^2)\,,
\end{equation}
so that the full model for the galaxy power spectrum in redshift space reads
\begin{align}
    T_{dw}(k)=T^2_{nw}(k)(1-e^{-k^2\Sigma_{\mathrm{pheno}}^2})+T^2(k)e^{-k^2\Sigma_{\mathrm{pheno}}^2}\,.
\end{align}

Where the $T_{nw}$ is the ``no-wiggle" matter transfer function obtained following the procedure of~\cite{Vlah:2015zda}, while $T$ is the matter transfer function.
In principle, $f^*(z)$ is a free parameter of the model. However, its value is completely unconstrained by the data, so we decide to fix it to its value in the $\Lambda$CDM universe with massless neutrinos, as in~\cite{McCarthy:2022qny}. We checked that this assumption does not impact the final results.
Note that, like in the EFT case, the galaxy power spectrum is defined with respect to the CDM+baryon power spectrum to avoid scale dependent biases (see previous section), so the transfer function that appears in Equation~\eqref{eq:model_ind} is the CDM + baryons transfer function. This transfer function is influenced by the presence of neutrinos, as they suppress the growth of the $cb$ perturbations on scales $k >   k_\mathrm{fs}$ Moreover, as in the EFT analysis,we assume that galaxies fall at the same rate as CDM and baryons, so that $\v_g=\v_{cb}$, as shown in~\cite{Verdiani:2025znc}. Let us stress again that the model for the power spectrum described in this section is completely agnostic about the underlying dark energy model or the particular parameterization of the dark energy equation of state.\\

\section{Observational effects}

\subsection{AP effect}
In the actual measurement of the galaxy power spectrum, we need to assume a cosmological model to convert the galaxy's position from angular and redshift coordinates into 3D comoving cartesian coordinates.
However, in general, it is possible that the fiducial model is different from the true cosmology. This can lead to a distortion in the observed galaxy power spectrum known as the Alcock-Paczynski (AP) effect~\cite{Alcock:1979mp}.
To account for this issue, we have to rewrite the true Fourier modes (denoted with $'$) in terms of the fiducial ones. This can be easily obtained from the Jacobian of the coordinate transformation between the real parallel and perpendicular distances $d_{\parallel},d_{\perp}$ and the fiducial ones.
We have
\begin{equation}
    k_{\perp}'=k_{\perp}/\alpha_\perp\,,\quad\quad k_{\parallel}'=k_{\parallel}/\alpha_\parallel\,,
\end{equation}
where
\begin{equation}
    \alpha_\perp=\frac{D_A(z)}{D_{A,\mathrm{fid}}(z)}\,,\quad\quad \alpha_{\parallel}=\frac{H_{\mathrm{fid}}(z)}{H(z)}\,.
\end{equation}
Thus, by defining $F\equiv\alpha_{\parallel}/\alpha_{\perp}$, one has
\begin{equation}
\begin{gathered}
    k'=\frac{k}{\alpha_\perp}\sqrt{1+\mu^2\left(\frac{1}{F}-1\right)}\,,\\
    \mu'=\frac{\mu}{F}\frac{1}{\sqrt{1+\mu^2\left(\frac{1}{F}-1\right)}}\,.
\end{gathered}
\end{equation}
The change of coordinates also rescales the overall multipoles by the determinant of the Jacobian, i.e., $\frac{1}{\alpha_\perp^2\alpha_\parallel}$. Note that, while in principle we are analyzing galaxy mock catalogs and the true underlying cosmology is known, the AP effect must be included in any case to consistently account for the different values of the cosmological parameters explored during the MCMC.
The final power spectrum multipoles are then given by
\begin{equation}
    P_{\ell,g}(k,z)=\frac{2\ell+1}{2\alpha_\perp^2\alpha_\parallel}\int d\mu L_{\ell}(\mu)[P_\mathrm{g}(k'(k),\mu'(\mu),z)]\,,
\end{equation}
In what follows, we divide the total observed lightcone into 6 redshift bins, evaluating our model at the effective redshift of that specific slice given by (see e.g.,~\cite{BOSS:2016psr})
\be
z_{\mathrm{eff}}=\frac{\sum_i^{N gal}w_{FKP,i}z_i}{\sum_i^{N gal}w_{FKP,i}}
\ee
where $w_{FKP}$ are the FKP weights. The specific binning scheme and the effective redshifts are shown in Tab.~\ref{tab:redshift} as well as the comoving volume and the mean number density of galaxies in each bin.

\begin{table}
    \centering
    \begin{tabular}{|c|c|c|l|l|}\hline
         $z_{\mathrm{min}}$&  $z_{\mathrm{max}}$ & $z_{\mathrm{eff}}$ & $\bar{n}$($h/$Mpc)$^3$&V(Mpc$/h$)$^3$\\\hline
         0.5&  0.8 & 0.664 & $4.61\times10^{-3}$&$12.86\times10^{8}$\\\hline
         0.8&  1 & 0.9 & $2.94\times10^{-3}$&$12.35\times10^{8}$\\\hline
         1&  1.2 & 1.1 & $2.00\times10^{-3}$&$14.75\times10^{8}$\\\hline
         1.2&  1.4 & 1.3 & $1.33\times10^{-3}$&$16.60\times10^{8}$\\\hline
         1.4&  1.6 & 1.5 & $8.94\times10^{-4}$&$17.96\times10^{8}$\\\hline
         1.6&  2& 1.79 & $4.79\times10^{-4}$&$38.44\times10^{8}$\\ \hline
    \end{tabular}
    \caption{Binning scheme adopted in this work. Mean number densities and comoving volumes are also shown for each redshift bin.}
    \label{tab:redshift}
\end{table}
We will employ the effective redshift approximation for the following reasons. The full 3D clustering cannot be correctly analyzed in Fourier space since, if we drop the flat sky and plane parallel assumption, the mapping between real and redshift space introduces mode coupling effects that make the standard power spectrum itself mathematically ill-defined~\cite{Zaroubi:1993qt}. One way to approximately account for this effect is the so-called unequal time formalism introduced in~\cite{RVa,RVb,Spezzati:2025ntb}. However, at leading order, unequal time effects are completely negligible when  a single tracer of the underlying dark matter field is analyzed.
\subsection{Window function}
Another observational effect that we must include in our analysis is the window function, which accounts for the angular and radial mask of the survey. This is done directly in Fourier space in order to avoid multiple Fourier transforms of the power spectrum model, which can be numerically unstable for very high $k$ due to the presence of the counterterms~\cite{DAmico:2019fhj,Beutler:2021eqq}.
In configuration space, the windowed multipoles of the correlation function are simply obtained by $\xi_{\ell,W}(s)=Q_{\ell,\ell'}\xi_{\ell'}(s)$, where $Q_{\ell,\ell'}=C_{\ell,\ell',\ell''}Q_{\ell''}(s)$, $Q_\ell(s)$ are the window function multipoles and:
\begin{align}
C_{0,\ell',\ell''} &=
\begin{bmatrix}
1 & 0  \\
0 & \frac{1}{5}  \\
\end{bmatrix}_{\ell',\ell''}\,, \quad
C_{2,\ell',\ell''} =
\begin{bmatrix}
0 & 1  \\
1 & \frac{2}{7}  \\
\end{bmatrix}_{\ell',\ell''}\,,
\end{align}
where repeated indices are summed over.
Going to Fourier space, we have that:
\begin{align}
    &P_{\ell,W}(k)=4\pi(-i)^\ell\int ds s^2j_\ell(ks)\xi_{\ell,W}(s)\\\nonumber
    &=\frac{2}{\pi}(-i)^\ell i^\ell\int ds s^2j_\ell(ks)\int dk' k'^2j_{\ell'}(k's)Q_{\ell,\ell'}(s)P_{\ell'}(k')\,,
\end{align}
now defining
\begin{equation}
    W_{\ell,\ell'}(k,k')=\frac{2}{\pi}(-i)^\ell i^\ell k'^2 \int ds s^2 j_{\ell}(ks)Q_{\ell,\ell'}(s)j_{\ell'}(k's)\,,
\end{equation}
we have the convolution.
\begin{equation}
    P_{\ell,W}(k)=\int dk W_{\ell,\ell'}(k,k')P_\ell'(k')\,.
\end{equation}
It can be shown that, given a galaxy catalog, the multipoles of the window function in configuration space can be estimated in the following way
\begin{align}
Q_\ell(s) &= i^{\ell} \int_0^\infty \frac{dk}{2\pi^2} \, k^2\VEV{P_{\mathrm{rand},\ell}(k)} j_\ell(ks)\,.
\end{align}
Namely, $Q_\ell(s)$ can be obtained by Fourier transforming the power spectrum of the random catalog needed for calculating the power spectrum itself. Crucially, the window function multipoles in Fourier space must have the same normalization as the power spectrum. see~\cite{Simon:2022adh,Beutler:2018vpe} for a detailed discussion.

\subsection{Integral constraints}
When constructing the galaxy overdensity field for power spectrum estimation, one must specify the mean galaxy density. In practice, this is taken to be the survey’s mean density, which ensures that the measured power spectrum vanishes at the survey scale—a correction known as the integral constraint. To prevent biases in cosmological parameter inference, the same constraint must also be imposed on the theoretical power spectrum model. While the integral constraint formally affects only the $k=0$ mode, the survey window function couples neighboring modes, causing its effect to extend to finite $k$.
The integral constraints-corrected power spectrum model reads~\cite{BOSS:2016psr, Beutler:2018vpe,Beutler:2021eqq,BOSS:2013uda}:
\begin{align}
    P_{\ell}(k)=&\int dk'k'^2W_{\ell,\ell'}(k,k')P_{\ell'}(k')\\\nonumber&-\frac{Q_\ell(k)}{Q_0(0)}\int dk'k'^2W_{\ell,\ell'}(0,k')P_{\ell'}(k')\,,
\end{align}
where $Q_\ell(k)=\VEV{P_{\mathrm{rand},\ell}(k)}$.
While we include the integral constraints correction in our model for completeness, since we use $k_{\mathrm{min}}=0.01h/\mathrm{Mpc}$, the effect of the integral constraint is negligible.

\section{Data Simulation and Analysis}

\subsection{Roman galaxy mock}

In this work, we analyze galaxy mocks on the lightcone, mimicking a population of H$\alpha$ galaxies expected to be observed by Roman. Such galaxies populate a cosmic volume of $2400$deg$^2$ at redshift $0.5<z<2$.

These mocks are obtained starting from the Last Journey gravity-only cosmological simulation~\cite{Heitmann:2020gux}: an extreme-scale N-body run carried out on the Mira supercomputer, where more than 1.2 trillion particles are evolved in a
($3.4$Mpc/$h$)$^3$ volume, assuming a fiducial cosmology consistent with Planck 2020~\cite{Planck2018}. %\red{(Yun: the input cosmological model of the mock needs to be given here.)}
The fiducial cosmological parameters of the Last Journey simulation are as follows: $\omega_b=0.0224,\omega_{\mathrm{cdm}}=0.1193,h=0.6766,\sigma_8=0.81,n_s=0.9665,w=-1$.

Substructure merger trees are defined based on the subhalo core tracking technique \cite{rangel_etal20, sultan_etal21}, and subhalo cores are placed into lightcone coordinates using the methodology described in \cite{Heitmann:2020gux}. Synthetic galaxies are grafted onto the merger trees using the Diffsky modeling framework: Diffsky is a computationally efficient model of the galaxy--halo connection based on parametrized physical prescriptions for galaxy star formation history \cite{alarcon22_diffstar, alarcon25_diffstarpop} and stellar population synthesis \cite{hearin_etal21_dsps}.

\begin{figure}
    \centering
    \includegraphics[width=0.85\linewidth]{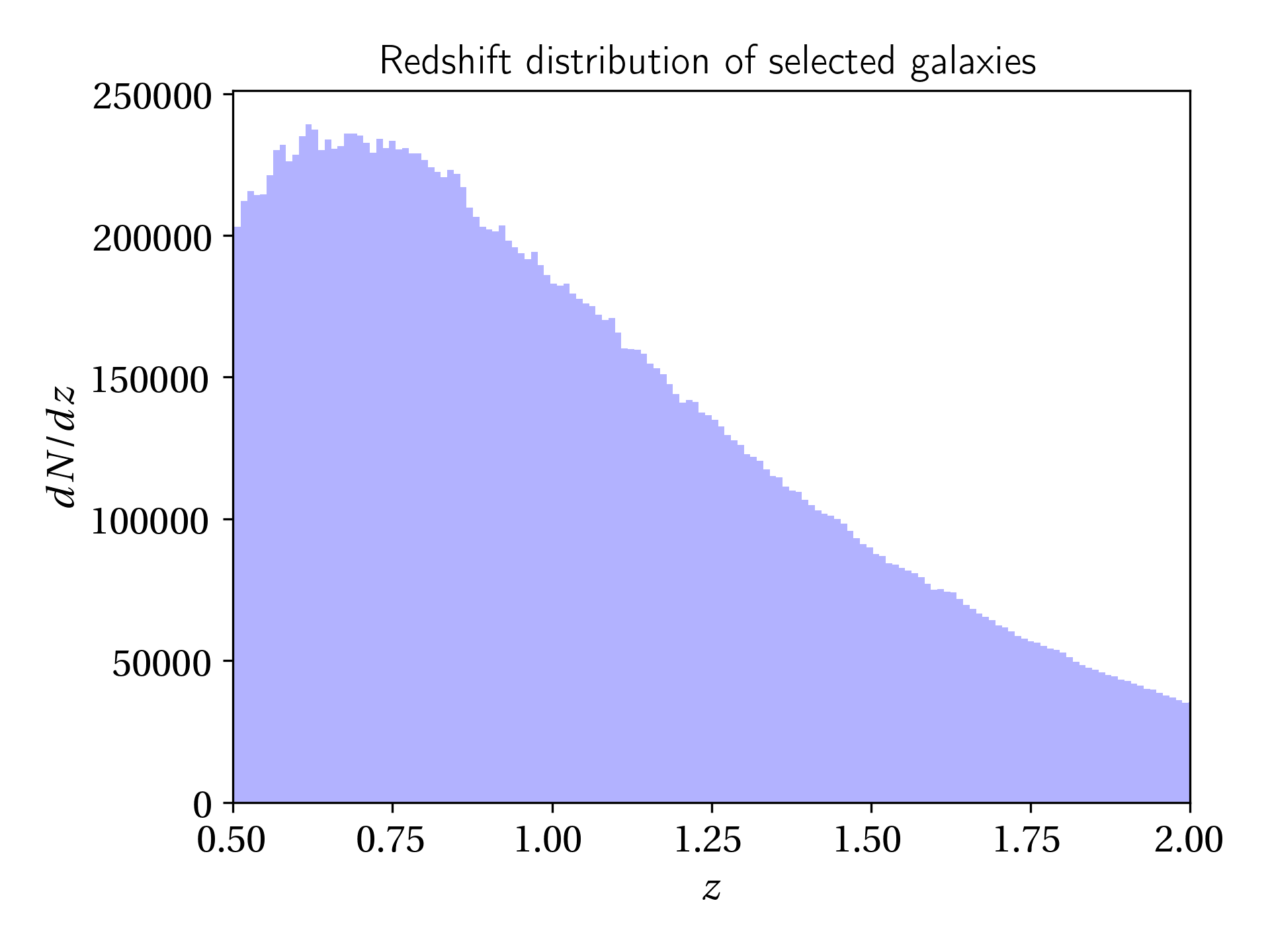}
    \caption{$dN/dz$ of the selected subsample of H$\alpha$ mock galaxies.
%    \red{(Yun: you need to make a new version of this figure with the y axis labeled.)}\FS{Done, let me know if it is ok.}
%    The number of galaxies is the integral over the curve.)}\FS{corrected}
}
    \label{fig:dndz}
\end{figure}

In order to select a galaxy sample resembling H$\alpha$ galaxies that will be observed by Roman, first we define $\epsilon_g=\mathrm{SFR}_g/{D_L^2}$ for each galaxy, where ${\rm SFR}_g$ is the star formation rate of the galaxy, and $D_L$ is the luminosity distance. We then rank-order galaxies by $\epsilon_g,$ and select our galaxy sample according to $\epsilon_g<\bar{\epsilon}$, where $\bar{\epsilon}$ is chosen so that the total number of galaxies is $N_{tot}\sim 2\times10^7$. This corresponds to a Roman galaxy sample of H$\alpha$ ELGs at $0.5<z<2$ with $f_{H\alpha}<1.5\times 10^{-16}$ers/s/cm$^2$, including only 20\% of H$\alpha$ ELGs at $0.5<z<1$ to ensure 2 emission lines are detected \cite{Wang:2021oec}.
%in line with current expectations for the Roman galaxy survey.
With this procedure, the redshift distribution of the selected sample is very similar to the one in ~\cite{Zhai:2019hjt} where galaxies with dust-attenuated H$\alpha$ flux higher than $10^{-16}$ erg s $^{-1}$ cm $^{-2}$ are selected.
The $dN/dz$ of the selected sample is shown in Fig.\ref{fig:dndz}.
We then apply RSD to the selected galaxies using the mapping~\cite{Kaiser1987,Hamilton:1997zq}:
\begin{equation}
    \mathbf{s}=\r+\frac{\v_r}{H(z)}\hat{r}\,,
\end{equation}
where $\v_r$ is the comoving peculiar velocity of each galaxy along the line of sight $\hat{r}$. No plane parallel approximation is used as $\hat{r}$ changes from galaxy to galaxy.

\subsection{Galaxy Power Spectrum Measurement}
For the measurement of the power spectrum multipoles, we use the public code \texttt{nbodykit}~\cite{Hand:2017pqn}. We used $512^3$ as the mesh grid size and the Triangular Shaped Cloud (TSC) algorithm for the mass assignment scheme.
We corrected for aliasing effects with the interlacing technique ~\cite{Jing:2004fq,Sefusatti:2015aex} and divided the obtained density field (in Fourier space) by the window function introduced during the mass assignment procedure ~\cite{Jing:2004fq}. We measured the power spectra in $k$-bins of width $\Delta k=0.01h/\mathrm{Mpc}$ from $0.01h/\mathrm{Mpc}<k<0.3h/\mathrm{Mpc}$.
The measured Poissonian shot noise is subtracted from the measured monopole of the galaxy power spectrum.

\subsection{Analytical Covariance Matrix}

\begin{figure*}[t]  % or [b] for bottom
    \centering
    \includegraphics[width=0.6\textwidth]{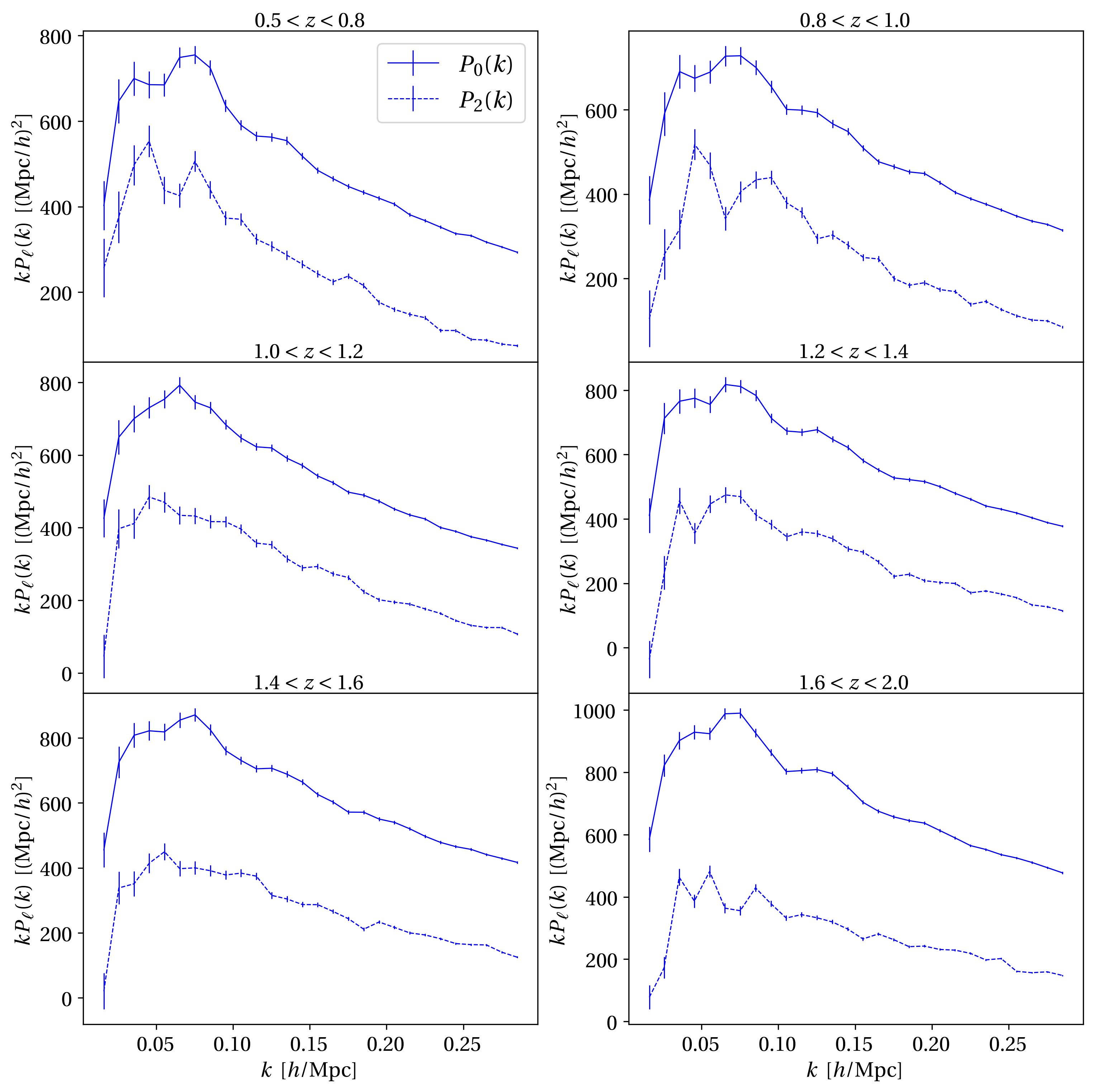}
    \caption{Measurements of the monopoles and quadrupoles of the galaxy power spectrum for the Roman mock, for the six redshift bins listed in Table.\ref{tab:redshift}.}
    \label{fig:data}
\end{figure*}

In our analysis, we employ an analytical Gaussian covariance matrix for the galaxy power spectrum multipoles. In each redshift bin, it is given by~\cite{Grieb:2015bia}:
\begin{align}\label{eq:cov}
    &\text{Cov}_{\ell,\ell'}(k,z_{\mathrm{eff,i}})=2\frac{(2\pi)^3}{V_iV_k}\frac{2\ell+1)(2\ell'+1)}{2}\\\nonumber&\times\int d\mu L_\ell(\mu)L_{\ell'}(\mu)\times\left(P_{g,\mathrm{lin}}(k,\mu,z_{\mathrm{eff,i}})+\frac{1}{\bar{n}_i}\right)^2\,,
\end{align}
where $V_k=4\pi k^2\Delta k$ and $\bar{n_i}$ are the mean number densities of galaxies in the $i$-th redshift bin.
The linear template for the galaxy power spectrum in the covariance is obtained by fitting the data while keeping the cosmology fixed at its fiducial values.
In principle, non-Gaussian contributions change the covariance error budget by $O(10\%)$ at the scales of interest, but they are subdominant compared to the errors arising from the marginalization on all the nuisance parameters, making their net impact on the final posteriors of cosmological parameters negligible\cite{Wadekar:2019rdu,Chudaykin:2020hbf}.
In Fig.~\ref{fig:data} we show the measured monopoles and quadrupoles of the galaxy power spectrum for the Roman mocks in the six redshift bins considered in this work. Errors are taken from the diagonal of the covariance in Equation ~\eqref{eq:cov}.
The inclusion of the hexadecapole has a negligible impact on the final cosmological constraints (due to its low SNR) while increasing the impact of prior volume effects (since it requires additional nuisance parameters that are poorly constrained~\cite{DESI:2024jxi}). For these reasons, we do not include the hexadecapole in our analysis.\\

\section{Neutrino mass constraints}
For the $\Lambda$CDM analysis, we vary the following cosmological and nuisance parameters.
\begin{align}
    &\{\omega_b,\omega_{\mathrm{cdm}}, h,n_s,\ln(10^{10}A_s)\} \cup\{b_{1,i},b_{2,i},b_{\mathcal{G}_2,i},b_{\Gamma_3,i}\}\\\nonumber &\cup\{c_{0,i},c_{2,i},\epsilon_{0,i},\epsilon_{2,i}\}
\end{align}
where $i=1,2....6$ is the redshift bin index.
We put wide flat priors on cosmological parameters and use a uniform prior $\mathcal{U}(0,1\mathrm{eV})$ for the sum of neutrino masses.
We utilize the following choice of priors for the nuisance parameters
\begin{align}
    &\mathcal{P}(b_{1,i})=\mathcal{N}(\bar{b}_{1,i},2)\,,\\\nonumber
   & \mathcal{P}(b_{2,i})=\mathcal{P}(b_{\mathcal{G}_2},i)=\mathcal{N}(0,2),
    \mathcal{P}(b_{\Gamma_3,i})=\mathcal{N}(\bar{b}_{\Gamma_3,i}2)\,,\\\nonumber &\mathcal{P}(c_{0,i})=\mathcal{N}(0,30),\mathcal{P}(c_{2,i})=\mathcal{N}(30,30)\,,\\\nonumber
   & \mathcal{P}(\epsilon_{0,i})=\mathcal{P}(\epsilon_{2,i})=\mathcal{N}(0,2)\nonumber\,,
\end{align}
where $\bar{b}_{1,i}$ are the best fit values obtained by fitting the power spectrum while keeping the cosmology fixed (these values are also used for the covariance matrix), and $\bar{b}_{\Gamma_3,i}$ is the value of the third order bias given by the co-evolution relation~\cite{Desjacques:2016bnm,Abidi:2018eyd}:
\begin{equation}
    b_{\Gamma_3,i}=\frac{23}{42}(b_{1,i}-1)\,.
\end{equation}

Moreover, we put a BBN prior on $\omega_b$, namely $\mathcal{P}(\omega_b)=\mathcal{N}(\omega_{b,\mathrm{fid}},0.00055)$~\cite{Schoneberg:2024ifp}, and a broad prior on $n_s$, i.e., $\mathcal{P}(n_s)=\mathcal{N}(n_{s,\mathrm{fid}},0.042)$, which corresponds to the $10\times$ Planck 2018 error bar on the spectral index \cite{Planck2018}. The maximum wavenumber used in the $\Lambda$CDM baseline analysis is $k_{\mathrm{max}}=0.2h/$Mpc. We refer to this as the baseline analysis. These specifications are very similar to those of \cite{DESI:2024hhd,DESI:2024jxi} when neutrino mass constraints are derived from the full shape analysis of DESI DR1 data and represent the most up to date constraints obtained using the EFT model.
Both in the $\Lambda$CDM analysis and in the model independent one, we assume three degenerate neutrino mass states.
We also show results in which Planck 2018 priors are used on $\omega_b,\omega_{\mathrm{cdm}},n_s$. We implement the priors from \cite{Zhai:2019nad} but marginalize over CMB distance prior parameters
in a simplified and conservative approach.
%that are not varied in this work.

\begin{figure}
%    \centering
    \includegraphics[width=1\linewidth]{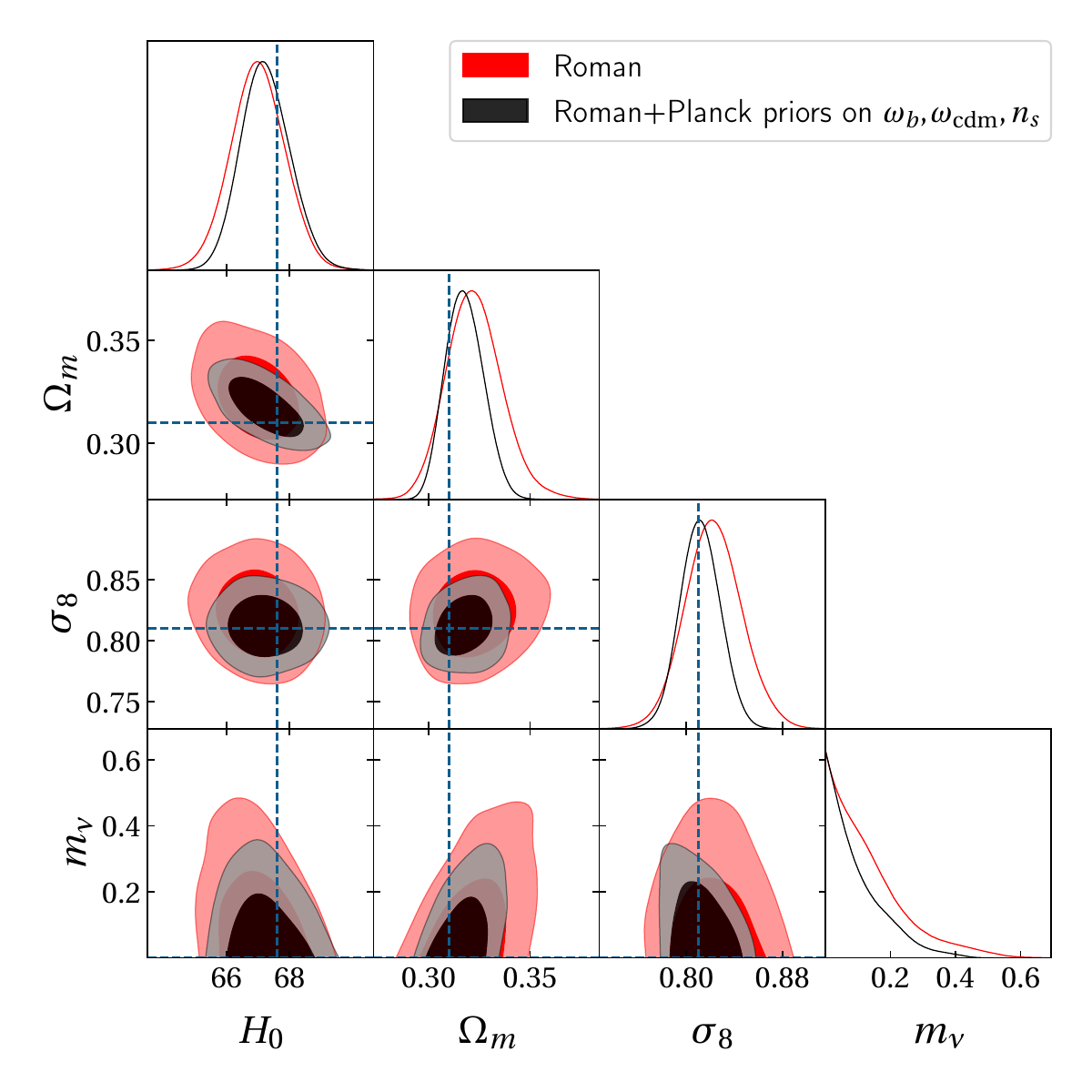}
    \caption{Posteriors of cosmological parameters obtained in the $\Lambda$CDM analaysis.
    The baseline analysis (red) uses the EFT model, with BBN prior on $\omega_b$ and $10\times$Planck prior on $n_s$(red). The results from adding Planck priors on $(\omega_b,\omega_\mathrm{cdm},n_s)$ to the baseline analysis (black). Both analyses used $k_\mathrm{max}=0.2h/$Mpc.}
%    \red{(Yun: the figure caption should be aligned on the left, and not centered!)}\FS{Done}

    \label{fig:EFT}
\end{figure}

\subsection{Neutrino Mass Constraints Assuming $\Lambda$CDM}

We present the results for the $\Lambda$CDM analysis in Fig. ~\ref{fig:EFT} for the following derived parameters, which are of the most interest for LSS analysis
\begin{equation}
    \{\Omega_m,\sigma_8,H_0,m_\nu\}\,,
\end{equation}
where $\Omega_m=(\omega_b+\omega_{\mathrm{cdm}}
+\omega_\nu)/h^2=\omega_m/h^2$, and  $\sigma_8$ is the variance of the linear matter overdensity field over a sphere of radius $8$Mpc/$h$ and $H_0=100h\,$km s$^{-1}$Mpc$^{-1}$. The fiducial cosmology is recovered within $1\sigma$ level, and constraints on the sum of the neutrino mass $m_\nu$ are as follows:
\begin{align}
   & m_\nu< 0.380\, (0.162), \quad95\%\,(68\%)\, \mathrm{C.L.}, \mathrm{\Lambda CDM}\\\nonumber &
   \hspace{0.4in} (\mathrm{BBN}+n_{s,10}+\mathrm{Roman\ GRS)}\,,\\\nonumber
   & m_\nu<0.276\,(0.121), \quad95\%\,(68\%)\, \mathrm{C.L.}, \mathrm{\Lambda CDM}\\\nonumber&
   \hspace{0.4in} (\mathrm{Planck\ priors\ on}\ \omega_b,\omega_{\mathrm{cdm}},n_s+\mathrm{Roman\ GRS})\,.
\end{align}
We can see that in our realistic forecast, if $\Lambda$CDM is assumed, the Roman data by itself could provide constraints on the sum of neutrino masses that are tighter than the current upper bound obtained from galaxy clustering data. In particular, we obtain neutrino mass constraints that are $\sim8\%$ tighter than the DESI DR1 EFT Full Shape $+$ DR1 BAO analysis in ~\cite{DESI:2024hhd} and $\sim42\%$ smaller than the $\Lambda$CDM constraints from the DESI DR1 Full shape Shape-Fit + DR2 BAO in \cite{Forero-Sanchez:2026bff}, using the same choice of priors on the baryon density, the spectral index, and the same range of scales.  Furthermore, our analysis shows that Roman could measure $H_0,\Omega_m,\sigma_8$ with $1.3\%,4.3\%,2.9\%$ precision at $68\%$ C.L, respectively, in line with DESI Full Shape measurements~\cite{DESI:2024jxi} and Euclid realistic forecasts~\cite{Euclid:2026cpu}.
When Planck priors on $\omega_b,\omega_\mathrm{cdm},n_s$ are added, the 68\% confidence interval for the sum of neutrino masses shrinks near $m_\nu=0.1\mathrm{eV}$, the threshold under which inverted hierarchy could be ruled out.

\subsection{Model-Independent Neutrino Mass Constraints}

For the model independent analysis, we consider the following free parameters:
\begin{equation}
    \{\omega_b,\omega_{\mathrm{cdm}},n_s,\bar{D}_{A,i},\bar{D}_{H,i},\log_{10}A_i,\beta_i\}\cup\{\sigma_{v,i},k^*_{i},Q_{i},C_{i},\alpha_{i}\}\,,
\end{equation}
where $\bar{D}_A$ and $\bar{D}_H$  are defined as
\begin{equation}
    \bar{D}_A(z_{\mathrm{eff},i})=\frac{D_A(z_{\mathrm{eff},i})}{D_{A,\mathrm{fid}}(z_{\mathrm{eff},i})},\quad \bar{D}_H(z_{\mathrm{eff},i})=\frac{H_{\mathrm{fid}}(z_{\mathrm{eff},i})}{H(z_{\mathrm{eff},i})}\,,
\end{equation}
and $\alpha_g$ is related to the growth of structure and is defined as
\begin{equation}
    \alpha_{g,i}=\frac{\sqrt{A_i}\beta_i}{f_{\Lambda CDM}(z_{\mathrm{eff},i})D_{\Lambda CDM}(z_{\mathrm{eff},i})P_{0,\Lambda CDM}}\,,
\end{equation}
where $f_{\Lambda CDM}$,$D_{\Lambda CDM}$ are the growth rate and growth factor in the fiducial $\Lambda$CDM model Universe, and $P_{0,\Lambda CDM}$ is the normalization of the matter power spectrum at $z=0$ in $\Lambda$CDM (for the fiducial model of the Last Journey simulation, its value is $P_{0,\Lambda CDM}=3.07\times10^6$). If the underlying cosmological model is indeed fiducial $\Lambda$CDM, then $\bar{D}_A,\bar{D}_H,\alpha_g=1$.
We have used the following priors
\begin{align}
   &\mathcal{P}(\bar{D}_{A,i})=\mathcal{P}(\bar{D}_{Z,i})=\mathcal{P}(\alpha_{g,i})=\mathcal{N}(1,0.3)\,,\\\nonumber
    &\mathcal{P}(\log_{10}A_i)=\mathcal{U}(6,7),\quad \mathcal{P}(\beta_i)=\mathcal{U}(0,1)\,,\\\nonumber
       & \mathcal{P}(\sigma_{v,i})=\mathcal{U}(0,10),\quad \mathcal{P}(k^*_i)=\mathcal{U}(0,1)\,,\\\nonumber
            &\mathcal{P}(Q_i)=\mathcal{P}(C_i)=\mathcal{N}(0,20),\quad \mathcal{P}(\alpha_i)=\mathcal{U}(0,1)\,.
\end{align}
We used Planck 2018 priors on $\omega_b,\omega_\mathrm{cdm},n_s$ from~\cite{Zhai:2019nad} (marginalized over $R$ and $l_a$) and $\mathcal{P}(m_\nu)=\mathcal{U}(0,1\mathrm{eV})$, as in the EFT analysis.

\begin{figure*}[t]
    \centering
    \includegraphics[width=1\linewidth]{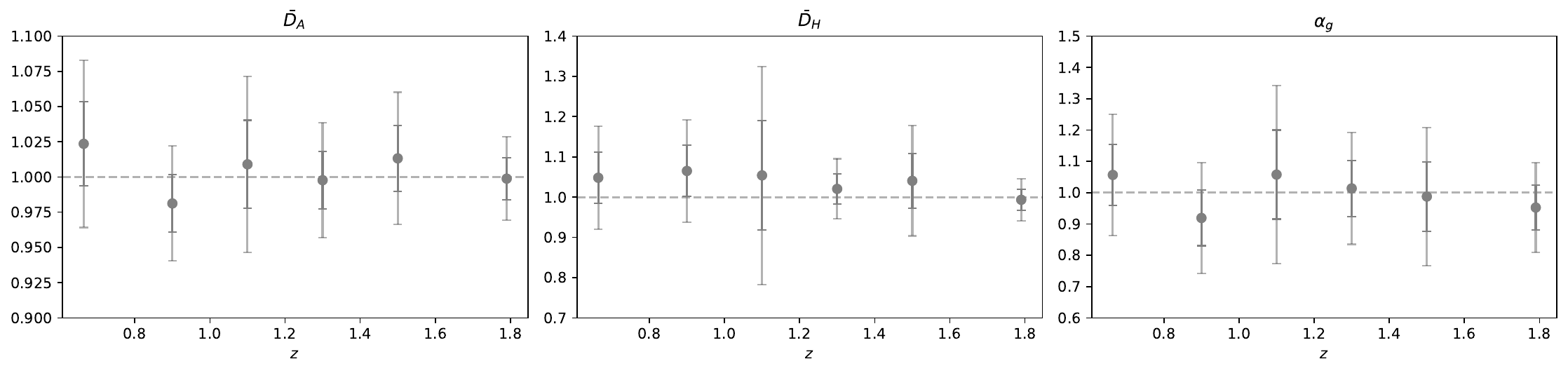}
    \caption{Posteriors means and $1\sigma$ (error bars) and $2\sigma$ errors (gray bands) for the $\bar{D}_A,\bar{D}_H,\alpha_g$ parameters for the six redshift bins.}
    \label{fig:alphas}
\end{figure*}

In the model independent analysis, the key parameters are:
\begin{equation}
    \{\bar{D}_{A,i},\bar{D}_{H,i},\alpha_g,\omega_m,m_\nu\}\,.
\end{equation}
Fig.\ref{fig:alphas} shows the mean of the posteriors of $\bar{D}_A,\bar{D}_H,\alpha_g$, along with the $1\sigma$ (error bars) and $2\sigma$ errors (gray bands) for the six redshift bins. We used the same $k_\mathrm{max}=0.2h/\mathrm{Mpc}$ as in the $\Lambda$CDM analysis, with the exception of the first redshift bin, where we used a lower $k_\mathrm{max}=0.17h/\mathrm{Mpc}$ to maintain the posteriors unbiased.
This is as expected since the linear regime increases with redshift, as perturbations grow in cosmic time.
We can see that we are able to recover the input cosmology within the $1\sigma$ level in all redshift bins for all three parameters.
%In Tab.~\ref{tab:alphas} we report the $1\sigma$ error on the $\bar{D}_A,\bar{D}_H,\alpha_g$ parameters for each redshift slice.
%\red{(Yun: This table is missing!)}

\begin{figure}
    \centering
    \includegraphics[width=0.6\linewidth]{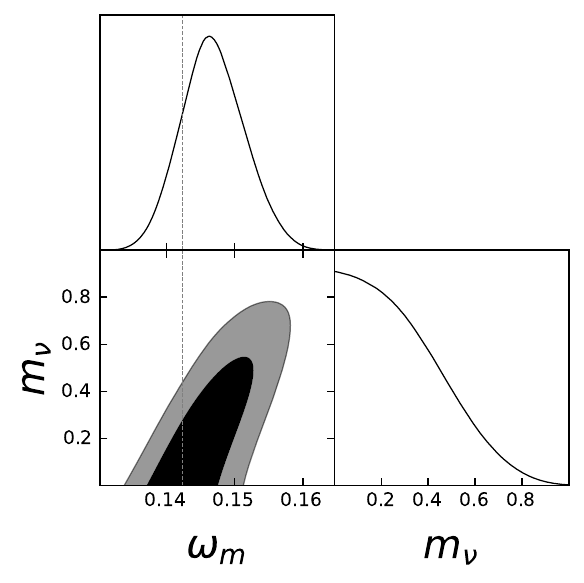}
    \caption{Cosmological parameters posteriors for the model independent analysis. Planck priors are used for $\omega_b,\omega_\mathrm{cdm},n_s$. }
    \label{fig:cosmo}
\end{figure}

We also plot the posteriors of the physical matter density $\omega_m$ and the sum of the neutrino masses $m_\nu$ in Fig.~\ref{fig:cosmo}. The model can recover the true cosmology within $1\sigma$ errors. In the model independent analysis, we report only constraints obtained when adding Planck priors to $\omega_b,\omega_\mathrm{cdm},n_s$, as the LSS data alone, when modeled using the model independent approach employed in this work, have low constraining power on the sum of neutrino masses. This happens because the effect of neutrinos on the growth of structure in the  model independent template is partially reabsorbed into the parameters $A_i$ and $\beta_i$ which are left free to vary in each redshift bin independently. This means that the model is sensitive to neutrino mass mainly at the level of the transfer function. Using Planck priors to break the degeneracy between $\omega_\nu$ and $\omega_b,\omega_\mathrm{cdm}$  in the expansion rate helps to increase this effect. The constraints on the sum of the neutrino mass are:

\begin{align}
   & m_\nu<0.63\,(0.36), \,\,95\%\,(68\%) \, \mathrm{C.L.}, \mathrm{model-independent}\nonumber\\
   & \hspace{0.2in}(\mathrm{Planck\ priors\ on}\ \omega_b,\omega_{\mathrm{cdm}},n_s+\mathrm{Roman GRS})\,.
\end{align}

\section{Discussion and Conclusions}

In this work, we have presented realistic forecasts for the constraining power of the Nancy Grace Roman Space Telescope on fundamental cosmological parameters, with particular emphasis on the absolute neutrino mass scale. To this end, we have analyzed simulated lightcone mock catalogs of H$\alpha$ galaxies spanning the redshift range $0.5 < z < 2$ over a sky area of $2400\ \mathrm{deg}^2$, mimicking the expected Roman High Latitude Wide Area Spectroscopic Survey. We performed a full-shape analysis of the galaxy power spectrum monopole and quadrupole in six redshift bins using two complementary theoretical frameworks: a model-dependent approach based on the EFT of LSS and a model-independent phenomenological method.

In the $\Lambda$CDM EFT analysis, we have shown that Roman alone could deliver neutrino mass constraints at the level of $m_\nu < 0.380\ \mathrm{eV}$ at 95\% C.L.\ ($m_\nu < 0.162\ \mathrm{eV}$ at 68\% C.L.) when combining BBN and a broad prior on $n_s$. These bounds are tighter than
%are already competitive with—and in fact exceed—
the current state-of-the-art constraints from galaxy clustering data
%, being $\sim\!8\%$ tighter than the DESI DR1 EFT Full Shape $+$ BAO analysis~\cite{DESI:2024hhd} and $\sim\!42\%$ tighter than the DESI DR1 ShapeFit $+$ DR2 BAO result~\cite{Forero-Sanchez:2026bff},
under identical prior choices and scale cuts. When external Planck priors on $\omega_b$, $\omega_\mathrm{cdm}$, and $n_s$ are incorporated, the 68\% upper bound tightens to $m_\nu < 0.121\ \mathrm{eV}$, approaching the threshold at which the inverted neutrino mass hierarchy could begin to be disfavored by cosmological data alone. In addition to neutrino mass, the Roman galaxy clustering data can constrain $H_0$, $\Omega_m$, and $\sigma_8$ with precisions of $1.3\%$, $4.3\%$, and $2.9\%$ at $68\%$ C.L., respectively, in line with Euclid realistic forecasts~\cite{Euclid:2026cpu} and DESI Full Shape measurements~\cite{DESI:2024jxi}.% \red{(Yun: fix this ref)}.

In the model-independent analysis, we have demonstrated that the phenomenological model can robustly provide unbiased measurements of the angular diameter distance $D_A(z)$, the Hubble parameter $H(z)$, and the growth of structure parameter $\alpha_g$ using similar scale cuts as the EFT model. When external Planck priors are imposed on the ($\omega_{cdm}, \omega_b, n_s$), the model-independent approach yields $m_\nu < 0.63\ \mathrm{eV}$ at 95\% C.I.\ ($m_\nu < 0.36\ \mathrm{eV}$ at 68\% C.I.). The weaker neutrino sensitivity compared to the $\Lambda\mathrm{CDM}$ analysis reflects the fact that, in the model-independent framework, the amplitude and growth rate are left free to vary independently in each redshift bin, allowing part of the neutrino suppression signal to be absorbed into unconstrained nuisance parameters. The sensitivity to neutrino mass in this approach is therefore driven primarily by the imprint of massive neutrinos on the shape of the transfer function, which is why external priors breaking the degeneracy between $\omega_\nu$ and the other density parameters are essential to extract meaningful bounds.

Together, these results highlight the complementary nature of the two approaches. The EFT-based $\Lambda\mathrm{CDM}$ pipeline maximizes statistical constraining power, but it requires the assumption of $\Lambda$CDM as the true cosmological model. The model-independent pipeline provides a robust, unbiased alternative that offers model agnostic measurements of geometric and growth observables and can be used for cosmological model selection without assuming a specific background evolution. The trade off for model independence is a weaker constraining power on $m_\nu$ compared to the $\Lambda$CDM analysis.

The constraints on the cosmological parameters presented here are expected to improve further when Roman full shape data are combined with Roman BAO measurements, as well as with complementary datasets from surveys such as DESI and Euclid.

\section*{Acknowledgments}
We thank Marco Marinucci, Alvise Raccanelli, Pierre Zhang and Zhongxu Zhai for useful discussions.
We gratefully acknowledge support from NASA Grant \#80NSSC24M0021, ``Project Infrastructure for the Roman Galaxy Redshift Survey'', and NASA ROSES Grant 12-EUCLID11-0004.

We gratefully acknowledge use of the Bebop and Improv supercomputers in the Laboratory Computing Resource Center at Argonne National Laboratory. Work done at Argonne was supported under the DOE contract DE-AC02-06CH11357. A portion of this work was supported by the OpenUniverse effort, which is funded by NASA under JPL Contract Task 70-711320, ‘Maximizing Science Exploitation of Simulated Cosmological Survey Data Across Surveys’.

\bibliography{biblio}

\end{document}